# Feature Integration Spaces: Joint Training Reveals Dual Encoding in Neural Network Representations

**Omar Claflin**

Independent Researcher
omarclaflin@gmail.com

**Abstract**

Current sparse autoencoder (SAE) approaches to neural network interpretability assume that activations can be decomposed through linear superposition into sparse, interpretable features. Despite high reconstruction fidelity, SAEs consistently fail to eliminate polysemanticity and exhibit pathological behavioral errors. We propose that neural networks encode information in two complementary spaces compressed into the same substrate: feature identity and feature integration. To test this dual encoding hypothesis, we develop sequential and joint-training architectures to capture identity and integration patterns simultaneously. Joint training achieves 41.3% reconstruction improvement and 51.6% reduction in KL divergence errors. This architecture spontaneously develops bimodal feature organization: low squared norm features contributing to integration pathways and the rest contributing directly to the residual. Small nonlinear components (3% of parameters) achieve 16.5% standalone improvements, demonstrating parameter-efficient capture of computational relationships crucial for behavior. Additionally, intervention experiments using 2×2 factorial stimulus designs demonstrated that integration features exhibit selective sensitivity to experimental manipulations and produce systematic behavioral effects on model outputs, including significant nonlinear interaction effects across semantic dimensions. This work provides systematic evidence for (1) dual encoding in neural representations, (2) meaningful nonlinearly encoded feature integrations, and (3) introduces an architectural paradigm shift from post-hoc feature analysis to integrated computational design, establishing foundations for next-generation SAEs.

**Code** — https://github.com/omarclaflin/LLM_Intrepretability_Integration_Features_v2
**Datasets** — WikiText-103 (publicly available)
**Extended version** — https://omarclaflin.com/llm-interpretability-project-dual-encoding-in-neural-network-representations/

## Introduction

**The Linear Superposition Assumption:** Current interpretability approaches fundamentally assume that neural network representations follow a linear superposition model (Elhage et al. 2022), where each activation can be decomposed into a sparse combination of interpretable features:

$$neural_activation = w_1 \times feature_1 + w_2 \times feature_2 + ...$$

This assumption underlies the success of Sparse Autoencoders (SAEs) (Bricken et al. 2023), which have demonstrated remarkable ability to discover interpretable features and achieve high reconstruction fidelity on neural activations. Within this framework, non-orthogonal feature representations are viewed as interference or compression artifacts—necessary evils that arise when neural networks attempt to represent more features than they have dimensions.

However, a fundamental puzzle remains: despite achieving high reconstruction fidelity, SAEs consistently fail to eliminate polysemantic features that respond to seemingly unrelated concepts (Bricken et al. 2023; Cunningham et al. 2023; Templeton et al. 2024; Chen et al. 2024). Recent work by Gurnee et al. has highlighted systematic limitations in sparse coding approaches. Examining logit distributions revealed that SAE reconstruction errors are pathological rather than random, and exhibit pathological behavioral errors when their reconstructions replace original activations, indicating missing computational structure.[1]

The mechanistic interpretability literature describes substantial "dark matter"—neural computation that remains unexplained even after extensive circuit analysis (Olah et al. 2024). These findings collectively suggest that the linear superposition assumption may be insufficient to capture the full complexity of neural representations.

[1] Gurnee, W. "SAE reconstruction errors are (empirically) pathological." AI Alignment Forum, March 29, 2024.https://www.alignmentforum.org/posts/rZPiuFxESMxCDHe4B/sae-reconstruction-errors-are-empirically-pathological

**Dual Encoding Hypothesis:** We propose that neural networks encode information in two complementary spaces that are compressed into the same neural substrate:

- **Feature Identity Space**: Represents what concepts are present in the input. This corresponds to the sparse features successfully captured by current SAE approaches—interpretable concepts like "Paris," "democracy," or "positive sentiment" that can be identified and measured independently.
- **Feature Integration Space**: Represents how concepts combine computationally to produce emergent meanings and behaviors. This may include the relationships between features that cannot be captured by linear combinations—the computational patterns that determine how "surprise" + "birthday" produces joy while "surprise" + "diagnosis" produces anxiety.

This dual encoding framework reframes non-orthogonal representations not as interference to be eliminated, but as computational structure encoding meaningful relationships between concepts. The persistent polysemanticity observed in neural networks may reflect the compression of both identity and integration information into the same representational space, rather than mere artifacts of insufficient capacity.

Additionally, this dual encoding hypothesis is distinct from existing analyses of feature relationships through co-activation patterns or similarity metrics, which capture statistical correlations between features across datasets. Instead, we focus on computational integrations—how features combine to produce emergent meanings that cannot be predicted from their individual activation patterns or co-occurrence statistics. While similarity analysis might reveal that "fire" and "hearth" (or "fire" and "forest") features often appear together, integration analysis reveals how their combination computes concepts like "warmth/comfort" (or "destruction/emergency") with behavioral consequences that emerge only from their joint activation.

Finally, traditional approaches attempt to address these limitations through post-hoc analysis—training SAEs first, then analyzing their failures. We propose an integrated architectural approach that captures both identity and integration patterns during training, preventing the systematic errors rather than detecting them after they occur.

**Neural Compression and Computational Structure:** Under this view, neural networks face a fundamental compression challenge: they must encode both the identity of relevant features and the computational relationships between them within limited representational capacity. A neuron that responds to both "late" and "party" (or "late" and "meeting") concepts may not simply be storing two unrelated features due to capacity constraints—it may be computing something about their relationship, such as "fashionable" (or "problematic").

This perspective offers a unified explanation for several puzzling phenomena in neural network interpretability: why polysemantic neurons are so robust across models, why certain feature combinations consistently appear together, and why interventions on individual features often produce complex, context-dependent effects, and why features exhibit sharp phase transitions during training as they crystallize from integration patterns into dedicated representations. If neurons encode computational relationships alongside feature identities, these observations become natural consequences of the underlying representational structure rather than obstacles to overcome.

To evaluate this dual encoding hypothesis, we explored a sequential training scheme, and then later developed a joint training architecture, that explicitly model feature identity and integration spaces. We investigate whether the improves reconstruction error, improves the pathology of logit probability distributions, whether they naturally organize into specialized computational roles when architectural constraints allow for separate feature identity and feature integration encoding, and whether nonlinear feature integrations can be confirmed with behavioral experimentation.

## Methods

**Experimental Pipeline:** Our methodology decomposes neural representations into two complementary encoding spaces through a three-stage pipeline:

1. sparse feature extraction via SAE training,
2. integration pattern capture using Neural Factorization Machine (NFM) predicting the SAE reconstruction residuals using the SAE features as inputs, and
3. integration feature analysis through a secondary SAE decomposition of the computed dense NFM embeddings, as a post-hoc interpretative step

```
Activations → Primary SAE → NFM → Secondary SAE
                  ↓           ↑  ↑              ↓
           Feature Identity → Residual error    ↓
                                                ↓
                  Interpretable Feature Integrations
                   (Secondary Feature Activations)
```

**Joint Training Architecture:** Following our initial exploration above, we also developed an integrated architecture that trains SAE and interactive components simultaneously in a single optimization phase, allowing natural specialization of feature types during learning.

All components were trained jointly using Adam optimizer (*lr=0.0001*, $\beta_1=0.9$, $\beta_2=0.999$) with linear learning rate decay from $1\times10^{-4}$ to $1\times10^{-5}$ over 80% of training steps using 5 million tokens from WikiText-103. Training proceeded in chunks of 10,000 tokens each, with 90/10 train-validation splits and batch size of 64 sequences. The TopK constraint (*1024, ~2.05% sparsity of 50k features*) provided automatic sparsity regularization, eliminating the need for additional sparsity loss terms beyond the reconstruction objective.

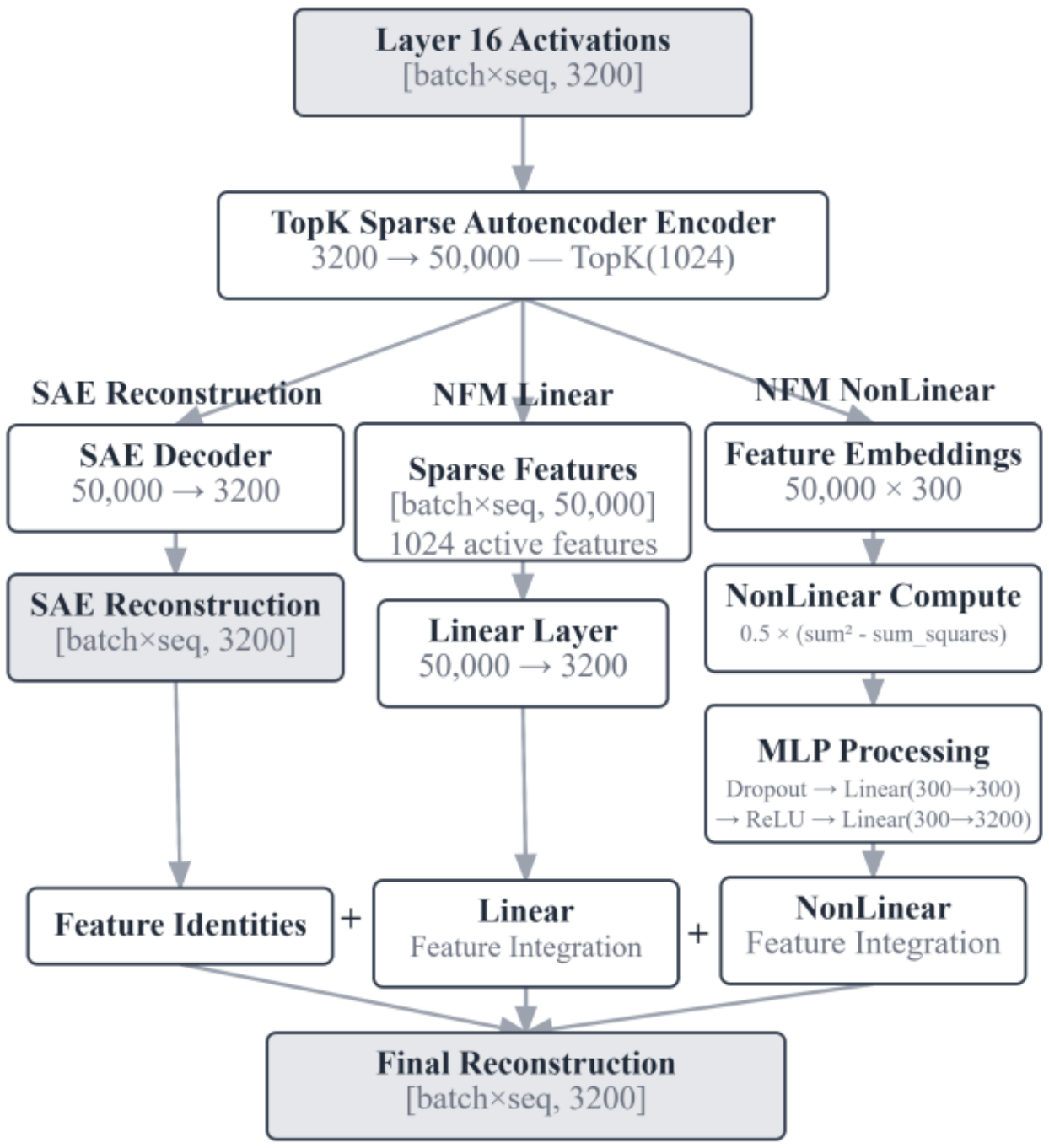

Figure 1. **Joint Architecture (SAE + NFM).** *Left side*: Traditional SAE pathway that decodes the SAE features to reconstruct the neural code. *Middle & Right Side*: New proposed components that integrate combinations of the SAE features, linearly and nonlinearly, to improve reconstruction of neural code while maintaining interpretability.

**Model and Data Configuration:** We conducted experiments using OpenLLaMA-3B with activations extracted from layer 16 (middle layer). The model was evaluated on WikiText-103, with tokenized sequences processed in 50-token windows. All experiments utilized a single NVIDIA RTX 3090 GPU with 24GB VRAM and 128GB system RAM.

**Stage 1: Sparse Autoencoder Training:** We trained a 50,000-feature TopK SAE achieving 0.136 reconstruction loss and 86.4% variance explained.

**Stage 2: Neural Factorization Machine Architecture:** NFMs capture feature integration patterns by predicting SAE reconstruction residuals:

$$\text{SAE error} = x_original - \text{SAE}(x_original)$$
$$\text{Residual Prediction (SAE error)} = \text{NFM(SAE features)}$$

Neural Factorization Machine combining linear and nonlinear integration components.

$$\text{Linear Output} = \Sigma_i\, w_i \times f_i + b$$
$$\text{Nonlinear Output} = 0.5 \times (\Sigma_i\, v_i f_i)^2 - \Sigma_i (v_i f_i)^2$$

where $v_i \in \mathbb{R}^k$ represents a learned embedding for feature i

The NFM was trained on 5 million tokens using Adam optimization (lr=1e-4) with K=300 embedding dimensions. This achieved 23.4% error reduction over the base SAE, with linear components contributing 95.5% and nonlinear contributing 4.5% of the improvement.

**Stage 3: Integration Space Analysis.** To analyze the computational structure captured by NFMs, we applied secondary TopK SAEs to the NFM intergration pathway, specifically targeting post-MLP1 vectors before ReLU activation. The secondary SAE used 25× expansion (300 → 7,500 features) with K=250 top active features of the primary SAE for each sample.

*Validation methodology*: We implemented 2×2 factorial stimulus designs (formal/informal × emotional/neutral) with systematic intervention experiments. Secondary SAE features were ranked by activation variance across experimental conditions, then subjected to clamping interventions at multiple levels (0×, ±1×, ±4×). Behavioral effects were measured through logit changes for category-relevant vocabulary sets.

**Stage 4: Experimental Validation Intervention testing**: We validated integration features through systematic clamping of both primary SAE features (via linear weight manipulation) and secondary SAE features (direct activation clamping). Effects were measured using logit differential analysis (ANOVA) across formality and emotion vocabulary categories to demonstrate specificity of the nonlinear behaviorial interaction effects of the integration features (secondary SAE).

**Controls**: For comparison against our NFM approach, a secondary SAE was trained directly on the original residuals showing no added significant reconstruction loss, compared to the NFM modelling approach. Additionally, other non-dead, but less active integration features were interrogated to demonstrate not all integration features showed statistically significant interactions in their behavioral effects. Finally, and not unsurprisingly, linear-component-only variants did not demonstrate nonlinear behaviorial interactions.

**Stage 5: Joint Training Implementation**: We trained a singular architecture which jointly trains SAE components and NFM integrations in a single loss function. The architecture combines three components in a residual manner:

$$final_reconstruction = sae_reconstruction + nfm_linear_out + nfm_nonlinear_out$$
$$total_loss = MSE(final_reconstruction - batch)$$

This approach allows primary SAE features to specialize naturally for either identity representation or integration computation during training, rather than retrofitting integration capture to pre-trained SAE features. For this particular implementation, we used a k-sparse approach (Gao et al. 2024) to provide automatic sparsity regularization for the SAE, which incidentally helps NFM stability, without requiring additional explicit loss terms, relying on MSE only.

**Stage 6: Evaluation of Joint Architecture.** In addition to reconstruction loss, and component contributions, we also looked at:

***Logit distribution:*** Analyzed predicted logits through KL divergence and cross-entropy loss by replacing model activations with architecture reconstructions and computing divergence from original model outputs, following Gurnee et al. methodology.

***Feature Orthogonality:*** We analyzed feature orthogonality through Gram matrix analysis (computing pairwise dot products between feature weight vectors to assess orthogonality patterns) and PCA analysis of feature weight distributions. We also examined the diagonal of the Gram matrix (looking at the squared norms).

***Bimodal Investigation:*** We analyzed features with different squared norms to investigate their differential contributions to direct reconstruction vs integration pathways.

***Parameter Impact****:* Finally, we looked at a component-wise analysis measuring reconstruction improvements and KL divergence improvements relative to parameter allocation across SAE, NFM linear, and NFM nonlinear components, using parameter count and total weight.

## Results

**Preliminary Quantitative Reconstruction Exploration.** The Neural Factorization Machine (NFM) approach achieved substantial improvements over sparse autoencoder baselines. Training on 5 million tokens, the combined SAE+NFM system demonstrated 23.18% error reduction on training data and 23.43% error reduction on validation data compared to SAE-only reconstruction which constrained our top K features to the top 250 features (*train error: 0.3813 → 0.2930*; *validation error: 0.3672 → 0.2811, [SAE only → SAE + NFM]*).

Component analysis revealed that linear combinations dominated the improvement, contributing 95.5% of the correction magnitude (linear: 0.2773, nonlinear: 0.0130), while higher-order (nonlinear) integration effects accounted for 4.5%. This suggests that NFMs capture both underspecified feature combinations that could be learnable by much larger SAEs and genuinely nonlinear integration patterns that may not be capturable through linear sparse coding approaches.

**Feature Specificity in Integration Space.** Using a stimulus-driven discovery approach, we identified primary SAE features responding to semantic dimensions of our experimental design: Feature 21607 (Emotion) and Feature 21781 (Formality) were selected based on maximal t-test differences across each stimulus conditions (a set of designed input stimuli containing our feature versus a stimuli set lacking that feature, generated by Gemini) directly.

Secondary SAE analysis on the NFM integration pathway revealed selective feature activation patterns. Among 7,500 secondary features, we identified features with distinct sensitivity profiles:

- *Feature 4022*: Highest ANOVA sensitivity (F=26.72, $p=2.85\times10^{-9}$) across experimental conditions
- *Feature 2020*: Highest activation in [formal,emotional] conditions (activation=0.517) but less interaction effects than 4022
- Counterexample features: *Feature 1113* showed no ANOVA sensitivity (F=0.022, p=0.996); *Feature 31* showed no activation differences across conditions

Distribution analysis across secondary features revealed a bimodal pattern: most features showed zero contribution to the primary feature dimensions, while a smaller subset exhibited normal distributions around meaningful contribution levels, with our target features appearing as outliers in the high-contribution tail.

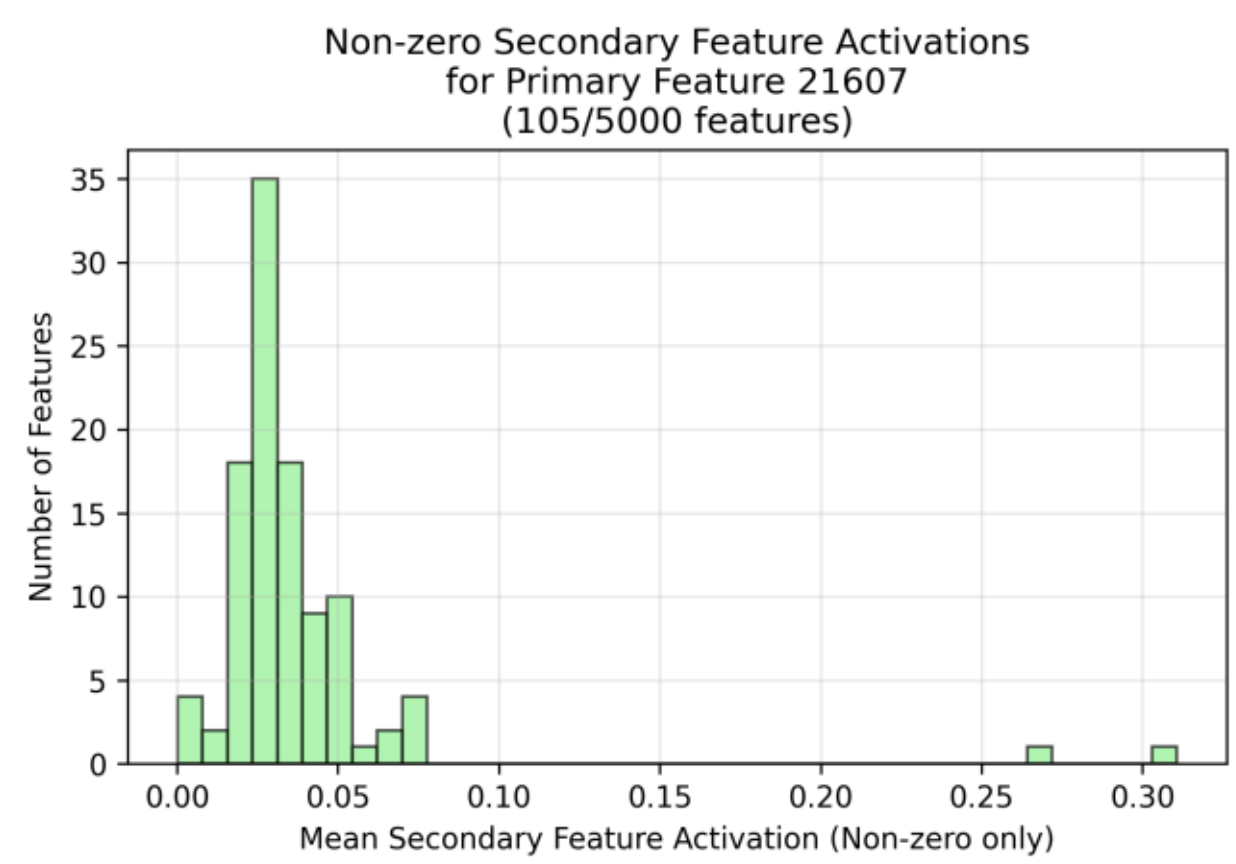

Figure 2. Distribution of secondary features (features from secondary SAE trained on NFM embedding) mean activation values when only primary feature (21607, Emotion) is naturally activated by our stimulus set. Only non-zero features are shown (vast majority are zero). *Note*: The other primary feature (21781, Formality, *not shown here*) produces a very similar secondary feature activation distribution.

**Intervention Validation and Behavioral Effects.** Systematic clamping experiments on *Feature 4022* demonstrated selective behavioral effects across vocabulary categories. Using clamping multipliers of [-4×, 0×, 4×], we measured logit changes for predetermined word sets (generated with Gemini 2.5):

- Formal/low-emotion: *"perhaps," "therefore," "consequently"*
- Formal/high-emotion: *"profoundly," "devastated," "extraordinary"*
- Casual/low-emotion: *"yeah," "basically," "whatever"*
- Casual/high-emotion: *"totally," "literally," "absolutely"*

Statistical validation confirmed significant interaction effects (F=5.06, p=0.027 for *formality*×*emotion* interaction), demonstrating that *Feature 4022* clamping produced non-additive effects across the 2×2 semantic space rather than simple main effects. Additionally, significant effects were seen on the other clamp values (-4.0: p=0.0273, 0.0 p=0.0257, 4.0 p = 0.0397), along with a differential impact of interaction effects by clamping (-4.0: 7e-4, 0.0: 1e-4, 4.0: 4e-4).

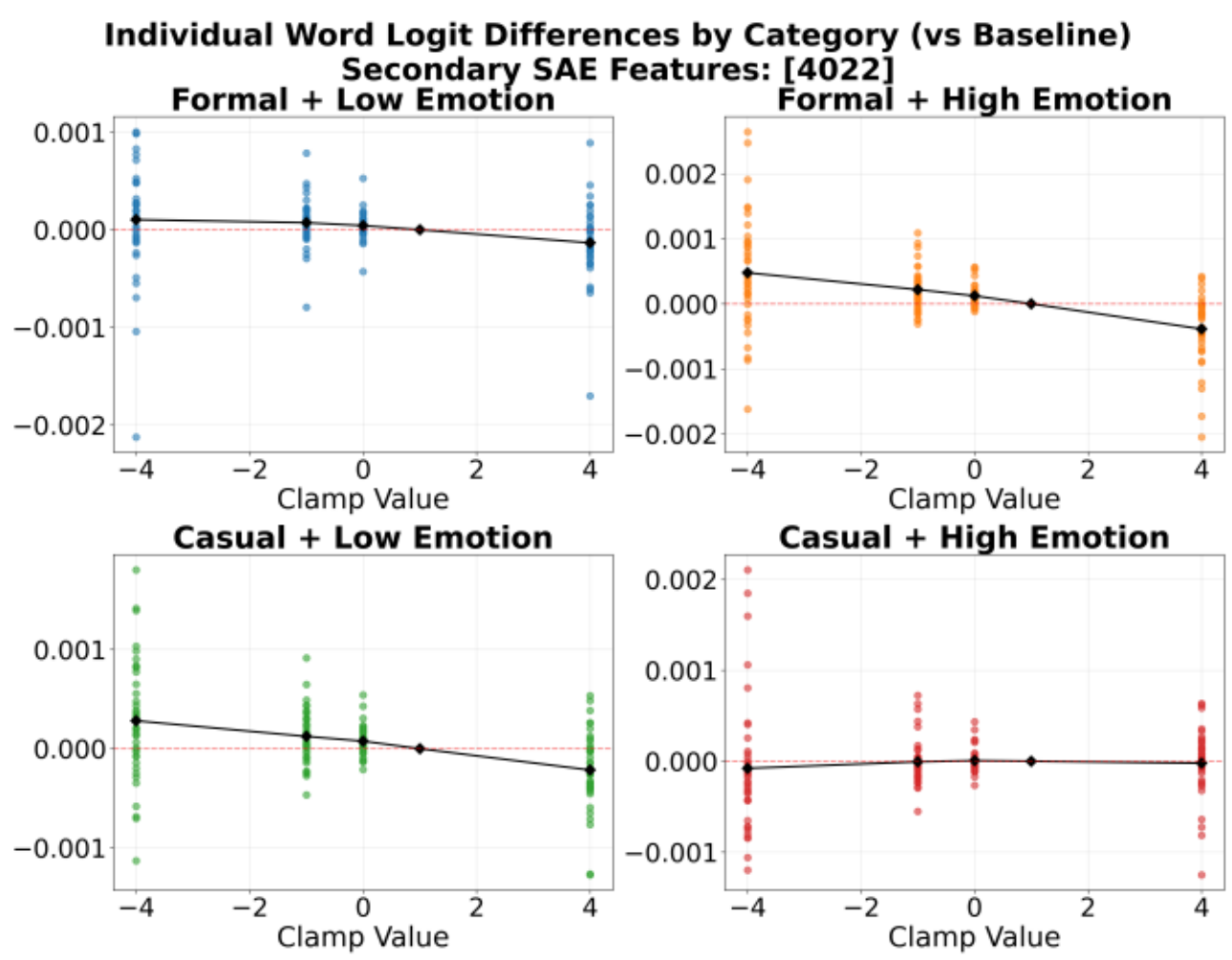

Figure 3. *X-axis*: Clamping values (-4, -1, 0, 1, 4) of integration feature, #4022. *Y-axis*: Reconstructed logit differences from baseline. *Four graphs*: 2x2 category of high/low for selected features (Emotion, Formality). Note the differential impact of the integration feature on bottom right (low Formality, high Emotion; reverse direction) driving the statistically significant ANOVA interaction, showcasing how the two features nonlinearly combine.

**Control experiments validated specificity:** This counter-factual exploration is not exhaustive but, along with the distribution plots, indicate some specificity of the nonlinear effects of the integration features discovered by our workflow.

- **Linear component clamping**: No ANOVA interaction effects observed from our linear layers, when comparing clamped linear integration weights vs baseline, in logits.
- **Non-sensitive feature clamping**: Feature 1113 showed no systematic patterns across categories (F>0.28, p>0.18, for all clamping ranges) despite significant activation.

**Sequential Architecture KL Divergence Analysis:** To further explore whether our improved reconstruction translates to better capturing distributions of logit fidelity compared to typical SAEs, we used Gurnee's test of KL divergence test vs a baseline test of a reconstruction with an $\varepsilon$-random error, reproducing a worse SAE divergence of 3x (compared to Gurnee's reported 2-4.5x range). Our new SAE-NFM sequential architecture showed *29.8% reduction* in pathological KL divergence errors across 17.8M measurements (SAE 2.99x, SAE+NFM 2.1x). Component-specific analysis revealed each of the components were sub-additive but significantly positive (*linear 29.8%, t=806.3; non-linear 0.9%, t=1375.8*). These results indicated our approach may address pathological structure in SAE reconstructions.

**Joint Training Architecture Performance:** Joint training substantially outperformed the sequential approach across reconstruction and behavioral metrics achieving 41.3% reconstruction improvement over the SAE (*joint model reconstruction error: 0.162, SAE reconstruction error: 0.275*), compared to 23% for sequential training.

Component analysis showed the NFM linear integration component comprised 94.0% of NFM parameters by mean absolute embedding weight (*0.347*), while nonlinear integrations comprised 6.0% (*0.022*), by the end of training.

**KL Divergence Error Analysis:** We validated the joint architecture against pathological KL divergence errors using 3.2M measurements. KL divergence by component showed 50.9%, 8.6%, and 51.9% (*t≥15.8*, *p<10e-6*) for the linear, nonlinear, and combination respectively. Cross-entropy loss showed a similar sub-additive pattern with linear, nonlinear, and combined components achieving 25.7%, 4.2%, and 26.2% improvements respectively (*t≥10.3*, *p<10e-6*). The nonlinear component alone achieved 16.5% of the total improvement using only 3% of the architecture's total parameters (1.1M, 9% of NFM integration parameters), demonstrating parameter efficiency in capturing nonlinear computational relationships.

**Feature Orthogonality and Emergent Specialization:** *PCA:* Principal component analysis showed both architectures achieved 90% variance explained with a similar number of principal components (~860), but the joint architecture required more dimensions across the first 50 components, indicating higher-dimensional feature representations.

*Gram matrix analysis:* Off-diagonal norm analysis revealed similar orthogonality of the joint architecture to the SAE. However, further analysis revealed the squared norms of our feature weights contained a distinct organizational pattern in the joint architecture versus an SAE (*below*).

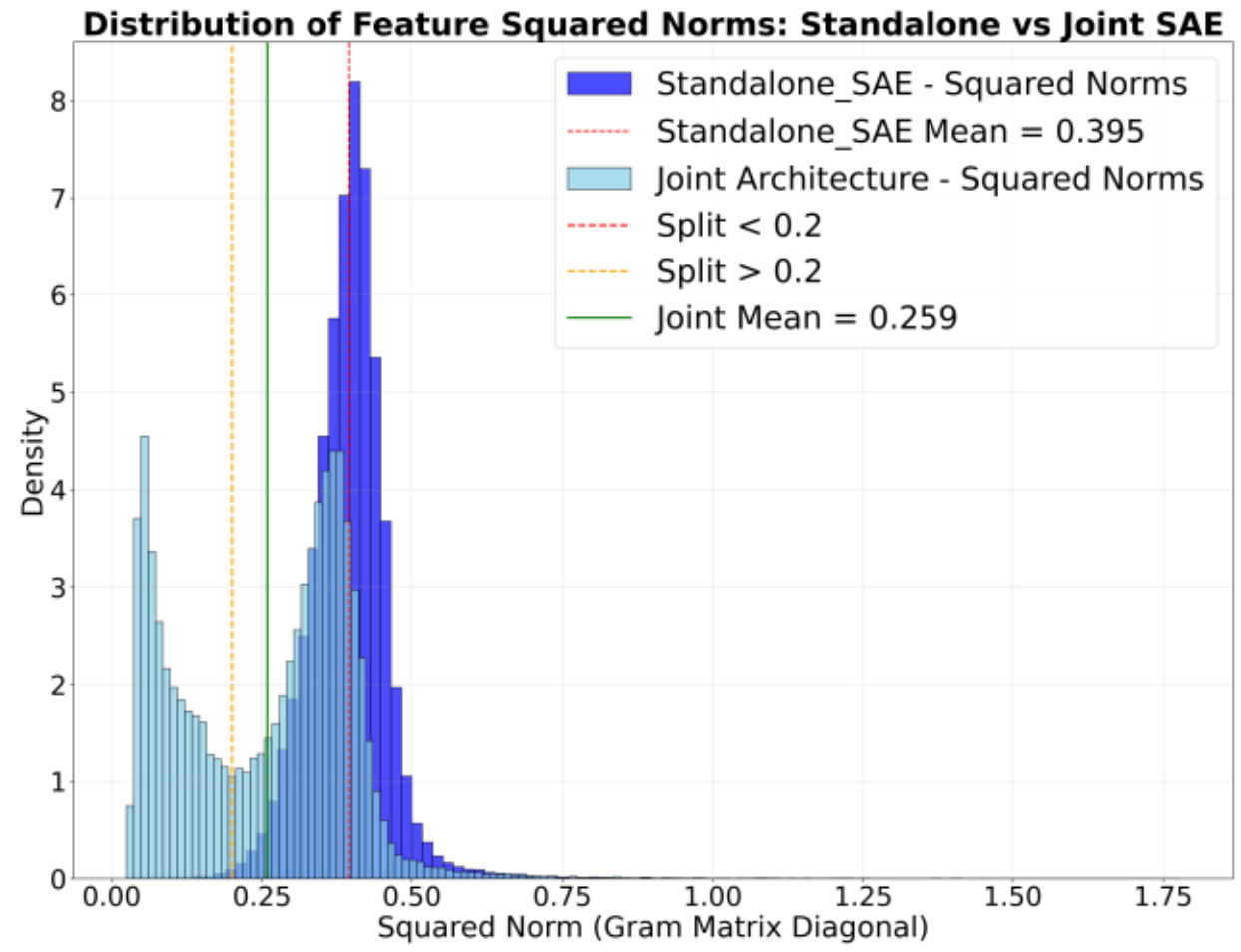

Figure 4. Distribution of diagonal squared norms of SAE features for a traditional SAE architecture (*blue*) compared to the distribution for our joint architecture (*turquoise*).

Gram analysis of our joint architecture showed a *bimodal distribution* with clear separation (split at 0.2) between low squared norm features (mean ~0.05) and moderate squared norm features (mean ~0.37). We follow this observation with another analysis looking at contributions to final reconstruction by weight, across all SAE features, for each component of the architecture.

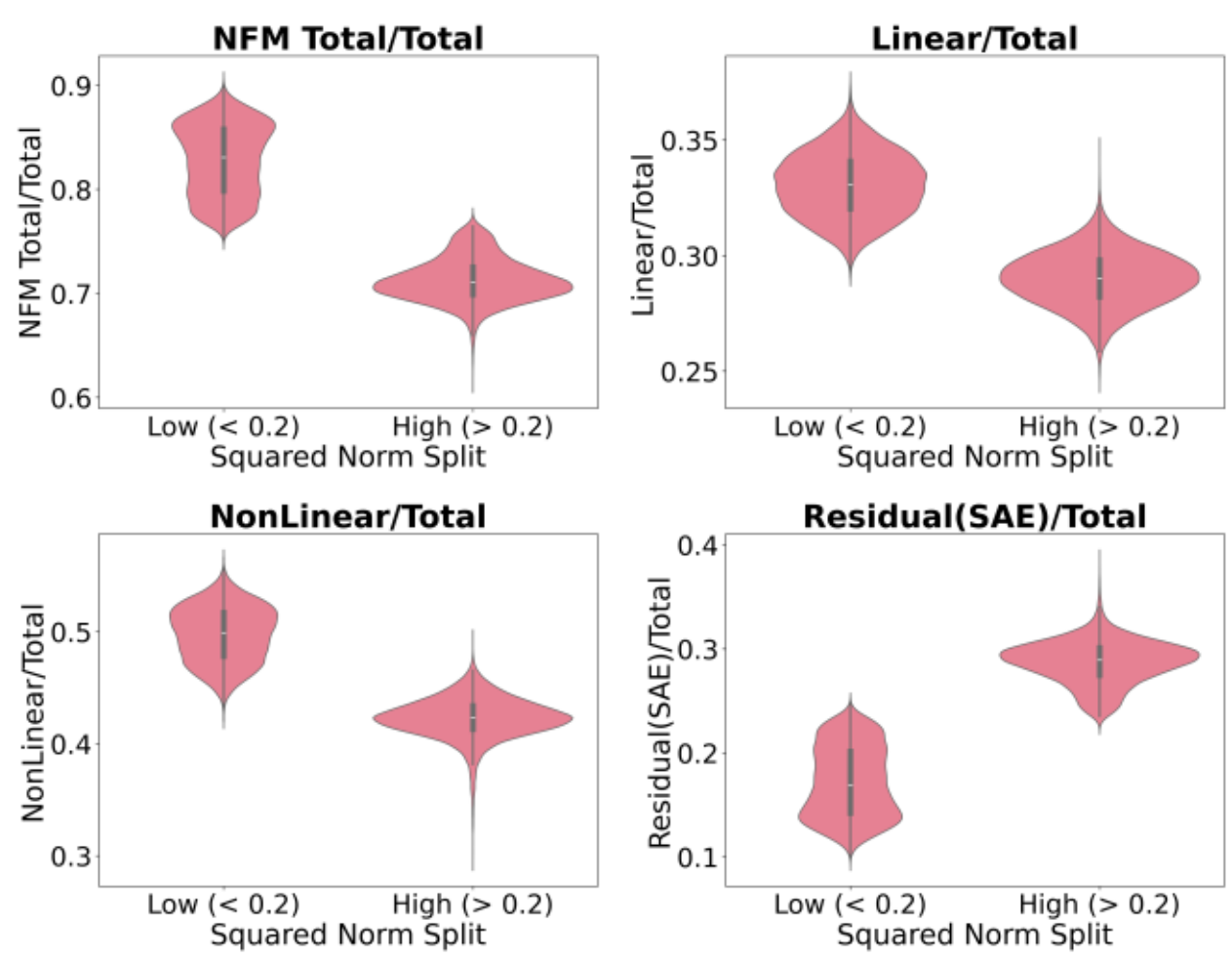

Figure 5. Violin plots of joint architectural contribution, split by the bimodal split (*low vs high squared norms, 0.2*), showing contributions of primary SAE feature to each component (*NFM, linear only, nonlinear only, residual*).

Overall, higher off-diagonal squared norms contributed more to the residual component directly (feature identity) while lower norm features contributed to the feature integrations (linear and nonlinear NFM components). Features with lower squared norms (<0.2) contributed significantly more to our integration components overall (*<0.2: 82.8% ,vs >0.2: 71.3%*) by mean absolute weight, and when broken down by the linear integration (*<0.2: 33.1% ,vs >0.2: 29.0%*), and nonlinear integration component (*<0.2: 49.7% ,vs >0.2: 42.3%*). While the opposite pattern of direct residual contributions from the SAE were more from our higher squared norm features (*<0.2: 17.2% ,vs >0.2: 28.7%*).

A simple correlative analysis, ignoring the bimodal split, reinforces this pattern with strong negative correlations (*r=-0.987, -0.867, -0.922, for total, linear, nonlinear*) between our squared norm and contributions by weight to our integration components. Only residual contributions coming from the SAE encoder show the opposite pattern (*r=0.987*).

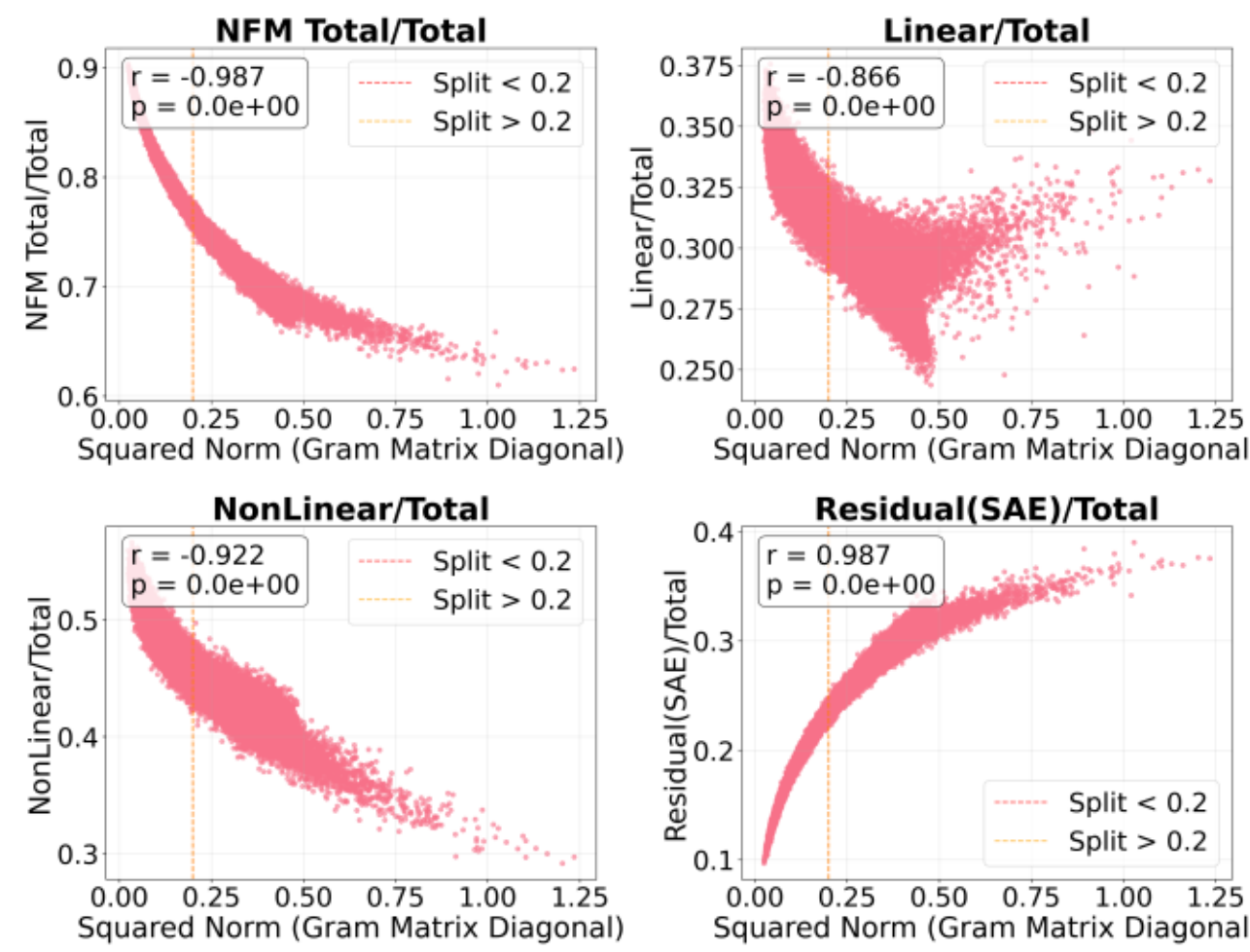

Figure 6. Contributions of each feature to each component of architecture by proportional weight. *X-axis*: Squared norms of features. *Y-axis*: Ratio of component weight over total weights. Each primary SAE feature contributes, with varying proportions, to all components of our architecture prior to neural activation reconstruction.

These findings suggest the joint architecture enables the SAE encoder to learn more diffuse, low squared norm feature identities that can be effectively utilized by downstream integration components, including their nonlinear combinations. In contrast, standard SAE training produces primarily high squared norm features that must simultaneously handle both feature identity and complex integrations within limited parameter space. This dual burden may cause standard SAEs to miss subtly distributed encodings throughout the layer.

## Discussion

**Implications for Neural Computation.** Our findings challenge the prevailing view of neural networks as sparse feature storage systems, revealing instead a dual encoding architecture where neural representations simultaneously compress both feature identity and feature integration information. The reconstruction improvement achieved by capturing integration patterns demonstrates that current sparse coding approaches, while successful at identifying interpretable features, systematically miss computational structure that is functionally significant for model behavior.

Polysemantic features may not represent compression artifacts to be eliminated, but rather computational units that encode relationships between concepts. The selective intervention effects we observed—where some features produce systematic 2×2 interaction patterns while others show no effects—suggest that polysemanticity may reflect meaningful computational roles rather than random interference patterns. This reframes the persistent polysemanticity observed even in high-capacity SAEs from a limitation to be overcome to evidence of fundamental computational organization.

The bimodal Gram matrix distribution provides direct empirical evidence for natural computational clustering. Joint training spontaneously develops two distinct feature populations: low squared norm features (mean=0.04) specializing in integration pathways, and moderately squared norm features (mean=0.37) contributing to direct reconstruction. This emergent organization validates our dual-encoding hypothesis at the representational level—the network naturally separates these computational roles when given architectural flexibility.

In contrast, standard SAE training shows unimodal squared norm distribution (mean=0.395), suggesting current approaches constrain all features to similar magnitude ranges. The strong negative correlations (r=-0.987) between squared norms and integration contributions reveal that low squared norm features specialize in computational relationships. This suggests that distributed, low-magnitude representations may be particularly suitable for integration precisely because their diffuse nature enables flexible recombination across integration components.

**Architectural Integration vs Post-Hoc Analysis.** Our comparison of sequential (23% improvement) and joint training (41.3% improvement) methods demonstrates a fundamental architectural principle: integration patterns benefit from simultaneous optimization with identity features rather than post-hoc capture of reconstruction residuals. This represents a paradigm shift from detecting missing computational structure to building architectures that naturally capture it during learning.

Traditional SAE approaches train sparse features first, then attempt to analyze or correct their limitations. Joint training allows the network to develop specialized feature types (naturally low squared norm features) that can be effectively recombined in integration components alongside higher squared norm features that handle direct reconstruction. This architectural flexibility eliminates the need to retrofit integration capture onto pre-trained sparse representations.

The 51.6% KL divergence reduction (vs 30% reduction in our sequential methodology) demonstrates that this architectural approach addresses fundamental limitations rather than providing incremental improvements. Joint training establishes a new standard for SAE architectures that integrate computational relationship modeling from the outset.

**Relation to Existing Work.** Our framework provides a possibly unifying explanation for several limitations identified in current interpretability research. The "dark matter" described in circuit analysis—computation that remains unexplained despite extensive feature identification (Lindsey et al. 2025)—may largely reflect missing integration patterns rather than inadequate feature discovery. Our demonstration that traditional maximum activation approaches fail to identify integration features, despite their clear functional effects, also may suggest that current interpretability methods may be systematically blind to this form of computation.

Our results provide a direct solution to the pathological KL divergence errors identified by Gurnee et al., achieving 51.6% reduction. This demonstrates that integration capture systematically addresses the behavioral limitations that have constrained SAE applicability—potentially providing the missing computational structure. The pathological nature of these errors reflects missing integration structure rather than random noise, and our architectural approach prevents these systematic failures rather than detecting them post-hoc. The systematic reconstruction errors in logit probability distributions find a natural explanation within our dual encoding framework: these errors reflect missing integration structure rather than random noise or capacity limitations.

Our work also addresses the "wrong abstraction level" problem frequently encountered in SAE research (Chanin, Shlegeris, and Brundage 2024; Makelov et al. 2024; Ayonrinde et al. 2024), where features appear either too specific or too general for interpretable analysis. Under our framework, these problems may reflect the imprecise fusion of identity and integration encoding: some apparent features may be integration patterns combined with features.

Related to this point, work on learning dynamics in toy models (Elhage et al. 2022), suggests discrete jumps of feature embeddings during training occur as the network learns to efficiently pack. Future learning dynamics work with our encoding hypothesis may reveal that some apparent integrations may simply be underspecified identity features awaiting sufficient encoding capacity or training data. Also, in an detailed footnote, theoretical efficiency advantages of nonlinear compression are dismissed based on neuron requirements (which seem to assume neurons cannot be reused polysemantically), the simple toy model (which ignores the nonlinearities of transformer architecture), and assumptions of nonlinearity in feature identity encoding. Our framework may provide a reasonable middle path to nonlinear interpretability, while still relying on linear interpretability in feature identity.

Unlike previous approaches examining SAE error suggesting systematic but unknown error sources (Engels, 2024), and static feature relationship methods that capture only co-occurrence patterns (Park et al. 2024), our feature integration approach reveals computational relationships—how features combine to produce emergent meanings which cannot be predicted from their individual activation patterns or statistical co-occurrence. This distinction is crucial for understanding the difference between features that merely appear together and features that compute together.

The parameter efficiency of our integration components (~32.3% of our total architecture's 496M parameters) contributing 40%+ gain in reconstruction loss suggests computational relationships require fundamentally different representational approaches than identity features. Interestingly, our *nonlinear* integration components (3% of parameters achieving 16.5% improvement) provide compelling evidence that nonlinearity in these integrations make substantial contributions to reconstruction performance.

Potentially, captured and encoded nonlinear relationships between encoded features, learned by the preceding nonlinear layers of the LLM serve a bigger role than previously thought. Overall, the nonlinear integration parameter efficiency indicates that while identity representation may require extensive sparse coding, integration patterns can be captured through targeted architectural components with disproportionate functional impact.

**Limitations and Future Work.** Scale constraints represent one primary limitation of this work. Our experiments on a 3B parameter model with 50k SAE features provide proof-of-concept evidence, but scaling to industrial-scale models with millions of features is likely challenging. The computational requirements of NFM training, while efficient compared to brute-force, scale super-linearly with feature count, may necessitate architectural innovations or more efficient approximation methods.

Integration interpretability presents ongoing challenges. While we demonstrated functional effects of integration features through systematic interventions, these features remain largely opaque to direct inspection. The failure of maximum activation analysis to yield interpretable patterns for integration features suggests need for specialized interpretability methods designed for computational rather than representational structure.

***Limitations of Traditional Feature Discovery.*** Discovery-oriented feature identification approaches are typically shown in interpretability experiments which has the advantage of being scalable and fairly objective, versus stimulus-oriented feature identification. Several discovery-oriented approaches were attempted, but we ran into issues with secondary feature identification. Relatively clean primary features could be identified in our primary SAE, along with corresponding secondary feature indices in which they contributed to, as well as demonstrating a behaviorial statistical interaction, but we ran into issues identifying what the secondary feature meant. This was in part because the behavioral output of our small Llama model was not reliable for traditional discovery-oriented approaches, even when using a primary SAE alone.

Maximum activation analysis on secondary features consistently returned conjunctive tokens ("that") or punctuation features (")") rather than interpretable semantic patterns. This failure occurred despite clear functional effects demonstrated through intervention experiments, suggesting that integration features may not correspond to simple activation maxima in natural text, likely encoding less obvious computational relationships.

These results provide converging evidence that neural networks encode feature integration patterns alongside feature identity, with integration features exhibiting selective sensitivity to experimental manipulations and producing systematic behavioral effects despite their opacity to traditional discovery methods.

These initial findings established proof-of-concept for dual encoding but revealed limitations in the sequential training approach. We therefore developed an integrated joint training architecture to test whether simultaneous optimization could improve both reconstruction fidelity and enable natural feature specialization (identity, integration). However, there are other possible architectures that may do this better or more efficiently.

The nonlinear integrations punch above their weight (3% of the total architecture parameters) but contribute non-additive gains that are largely overlapped by the linear integration gains (for both KL improvement and reconstruction error), potentially suggesting most feature computation is lower order. Also, in their current form, the nonlinear bi-interaction pooling approach in NFMs is efficient but conflates all higher-order feature combinations (2-way, 3-way, 4-way, etc.) in a compact dense parameter space. Other computationally efficient routes to explore nonlinear integrations without conflation may be advantageous.

Methodological extensions could address several current limitations: (1) Dynamic analysis of how integration patterns evolve during training could reveal the mechanisms by which computational relationships crystallize into identity features. (2) Cross-model validation could establish whether specific integration patterns represent universal computational primitives. (3) Cross-layer analysis could demonstrate the dynamics of feature integration as activity gets processed through layers. (4) Application to larger models could test whether the linear/nonlinear integration split observed here reflects fundamental properties of neural computation or artifacts of limited scale.

**Conclusion:** This work provides the first systematic evidence for dual-encoding spaces in neural network representations and introduces an architectural solution that achieves substantial improvements across multiple validation metrics. Joint training delivers 41.3% reconstruction improvement and 51.6% reduction in pathological KL divergence errors while spontaneously developing bimodal feature organization that validates our dual-encoding hypothesis. Critically, systematic intervention experiments revealed integration features with selective sensitivity to experimental manipulations, producing statistically significant behaviorial interaction effects (F=5.06, p=0.027) across semantic dimensions rather than simple main effects. The architecture demonstrates that computational relationships can be captured efficiently (32.3% of parameters achieving 41.3% improvement) when separated from identity representation.

This work establishes an architectural paradigm shift from post-hoc feature analysis to integrated computational design. Rather than training sparse autoencoders and then analyzing their limitations, joint training enables natural specialization where low squared norm features form distributed definitions that can be more effectively utilized by specialized integration components, while higher squared norm features must serve direct reconstruction. This emergent organization—evidenced through bimodal Gram matrix distributions and systematic specialization patterns—demonstrates that networks naturally separate identity and integration encoding when given appropriate architectural flexibility.

Methodological contributions include the first demonstration of separable feature identity and integration encoding, systematic approaches for detecting and intervention on the computational components that combine features, and stimulus-oriented validation methodologies that may have advantages for establishing functional significance beyond typical discovery-oriented approaches.

Broader implications extend beyond interpretability to fundamental questions about neural computation, AI safety, and the relationship between artificial and biological neural systems. Understanding how networks integrate information to produce emergent behaviors is crucial for developing reliable, controllable AI systems and for advancing theories of intelligence itself.

The convergence of reconstruction improvements, behavioral validation, and mechanistic understanding positions this approach as a foundation for next-generation sparse autoencoder architectures. By solving known limitations (pathological errors), providing architectural innovation (joint training), and revealing natural computational organization (emergent specialization), this work advances both the theoretical understanding and practical implementation of neural network interpretability.